# Absence of diode effect in chiral type-I superconductor NbGe$_2$

Dong Li[1,2,8]*, Zouyouwei Lu[1,3,8], Wenxin Cheng[1,3,8], Xiaofan Shi[1,3,8], Lihong Hu[1,3], Xiaoping Ma[1], Yue Liu[1,3], Yuki M. Itahashi[2], Takashi Shitaokoshi[2], Peiling Li[1], Hua Zhang[1], Ziyi Liu[1], Fanming Qu[1,3,4,6], Jie Shen[1,3,4], Qihong Chen[1,3]*, Kui Jin[1,3,4,5], Jinguang Cheng[1,3], Jens Hänisch[7], Huaixin Yang[1,3], Guangtong Liu[1,3,4,6]*, Li Lu[1,3,4,6], Xiaoli Dong[1,3,4,5], Yoshihiro Iwasa[2], and Jiangping Hu[1,3]

[1] *Beijing National Laboratory for Condensed Matter Physics, Institute of Physics, Chinese Academy of Sciences, Beijing 100190, China.*

[2] *RIKEN Center for Emergent Matter Science (CEMS), Wako 351-0198, Japan.*

[3] *School of Physical Sciences, University of Chinese Academy of Sciences, Beijing 100049, China.*

[4] *Songshan Lake Materials Laboratory, Dongguan, Guangdong 523808, China.*

[5] *Key Laboratory for Vacuum Physics, University of Chinese Academy of Sciences, Beijing 100049, China.*

[6] *Hefei National Laboratory, Hefei, Anhui 230088, China.*

[7] *Institute for Technical Physics, Karlsruhe Institute of Technology, 76344, Eggenstein-Leopoldshafen, Germany*

[8] These authors contributed equally: Dong Li, Wenxin Cheng, Zouyouwei Lu, Xiaofan Shi.

* Correspondence to: dong.li.hs@riken.jp, qihongchen@iphy.ac.cn, gtliu@iphy.ac.cn

**Symmetry elegantly governs the fundamental properties and derived functionalities of condensed matter. For instance, realizing the superconducting diode effect (SDE) demands breaking space-inversion and time-reversal symmetries simultaneously. Although the SDE is widely observed in various platforms, its underlying mechanism remains debated, particularly regarding the role of vortices. Here, we systematically investigate the nonreciprocal transport in the chiral type-I superconductor NbGe$_2$. Moreover, we induce type-II superconductivity with elevated superconducting critical temperature on the artificial surface by focused ion beam irradiation, enabling control over vortex dynamics in NbGe$_2$ devices. Strikingly, we observe negligible diode efficiency ($Q < 2\%$) at low magnetic fields, which rises significantly to $Q \sim 50\%$ at high magnetic fields, coinciding with an abrupt increase in vortex creep rate when the superconductivity of NbGe$_2$ bulk is suppressed. These results unambiguously highlight the critical role of vortex dynamics in the SDE, in addition to the established symmetry rules.**

**Introduction**

Stemming from the nonreciprocal phenomena in quantum materials [1-3], novel functionalities such as the superconducting diode effect (SDE) have been achieved in prototype devices through symmetry engineering [4]. Analogous to the semiconductor p-n junctions, superconducting diodes carry dissipationless supercurrent in one direction and show finite resistance in the opposite direction. An asymmetry parameter, called the diode efficiency $Q \equiv \frac{|I_c^+ - I_c^-|}{(I_c^+ + I_c^-)/2}$, was introduced to quantify the difference between the opposite critical currents $I_c^+$ and $I_c^-$ in SDE [5]. The SDE is crucial in realizing energy-efficient and quantum-based superconducting circuits, attracting extensive attention from the superconducting electronics community [6]. However, intense controversy has recently emerged regarding the mechanism of SDE [7]. Specifically, mainstream theories suggest that the SDE arises from the intrinsic inequality of Cooper-depairing currents along opposite directions [8-11]. In contrast, the experimentally measured $I_c$ is considered by other groups to be the vortex-depinning current from a practical viewpoint [12] and the SDE is hence thought to originate from the asymmetric vortex pinning potential [7,13-15].

The central question is how to avoid or even regulate the vortex effect in the study of SDE. Vortices are inevitably generated by the negative interface energy in type-II superconductors but are absent in type-I superconductors with positive interface energy [16]. On the other hand, as one kind of nonreciprocal response, the superconducting diodes require intrinsic space-inversion symmetry breaking. Therefore, investigating SDE in chiral type-I superconductors is an elegant method to circumvent the complicated vortex effect. However, both type-I superconductivity and chiral structures are unusual, rendering the overlap extremely rare. The only identified candidates for chiral type-I superconductors are polycrystalline BeAu (space group $P2_13$) [17] and single crystalline NbGe$_2$ (space group $P6_222/P6_422$) [18] so far. NbGe$_2$ belongs to the Kramers-Weyl semimetals [19] with a noticeable antisymmetric spin-orbit coupling (SOC) effect [20]. Meanwhile, the chiral symmetry facilitates the formation of a coupled electron-phonon liquid in NbGe$_2$ below $T' \sim 50$ K [21]. The electron fluid property leads to an ultralow resistivity in NbGe$_2$ crystals, posing a large obstacle to comprehensively investigating the subtle nonreciprocal charge transport [22,23].

This challenge can be overcome by preparing micro-sized devices using the well-established focused ion beam (FIB) technique [24]. FIB offers an excellent method for investigating quantum materials in the mesoscopic regime, particularly for materials like NbGe$_2$ that are incompatible with mechanical exfoliation. Another advantage of FIB devices is their flexibility in tailoring geometry, enabling diverse experimental configurations for transport measurements. Here, we successfully fabricated the NbGe$_2$ devices using FIB and systematically investigated their nonreciprocal transport. An unexpected outcome of FIB milling is the superconducting amorphous NbGe surface, which coincidentally exhibits type-II superconductivity with a superconducting critical temperature $T_c$ higher than that of NbGe$_2$ bulk. This artificial structure allows for the incorporation of vortex dynamics into NbGe$_2$ devices by suppressing the bulk superconductivity. While most nonreciprocal transport results align well with theoretical predictions, the field dependence of diode efficiency $Q$ is in strong contrast to anticipations. The main message of this paper is that the SDE is negligible ($Q < 2\%$) at low field regimes but emerges ($Q$ exceeding 50%) at high field regimes, corresponding to the strongly suppressed or activated vortex dynamics. Our study reveals that, in addition to the well-known symmetry rules, the ubiquitous vortex effect plays a nonnegligible and even critical role in realizing the SDE.

**Results and Discussion**

**Type-I superconductivity in chiral NbGe$_2$**

NbGe$_2$ belongs to the hexagonal C40 structural group, with right- and left-handed helix structures along the *c*-axis corresponding to space groups $P6_222$ (No. 180) and $P6_422$ (No. 181), respectively (Fig. 1a). The left-handedness of these two enantiomers in the studied NbGe$_2$ crystal was confirmed using single-crystal diffraction (Supplementary Table 1). Superconductivity in NbGe$_2$ was first found in 1978 [25] and has recently been revisited due to its chiral crystal structure [18,20,21,26,27]. Fig. 1b shows the temperature-dependent resistivity of a NbGe2 crystal with a residual resistivity ratio RRR = 96. The $T_c$ is consistent across electrical transport ($T_c$ = 1.96 K) and magnetic susceptibility measurements ($T_c$ = 2.01 K). It is worth noting that the magnetic susceptibility is identical in both the zero-field-cooled (ZFC) and field-cooled (FC) measurements (Fig. 1c). In type-II superconductors, the diamagnetism measured in the FC mode is smaller than that of the ZFC mode due to the trapped vortices. However, it has been confirmed that the superconductivity in NbGe$_2$ exhibits rare type-I behavior [18,20], which is responsible for the absence of the vortex pinning effect in our FC measurement. As the temperature decreases from 1.8 K to 400 mK (Fig. 1d), we observe a crossover in NbGe$_2$ from type-I superconductivity at low fields to type-II/1 superconductivity at higher fields [28]. The so-called type-II/1 superconductivity significantly differs from the conventional type-II superconductivity, although magnetic flux will penetrate both superconductors. The interaction between flux quanta is attractive, rather than repulsive, in type-II/1 superconductors [29]. This feature strongly suppresses the vortex dynamics in type-II/1 superconductivity, rendering it more like type-I than the conventional type-II superconductivity.

To enhance the charge transport signals, we fabricated micro-sized NbGe$_2$ devices using FIB (Methods). As shown in Fig. 2a, we have deliberately designed the NbGe$_2$ device D3 to simultaneously measure the *c*-axis and *a*-axis transport properties. The *a*-axis resistivity is higher than the *c*-axis resistivity in NbGe$_2$, consistent with previous report [21]. As the temperature-dependent resistivity remains consistent between the NbGe$_2$ bulk and NbGe$_2$ devices in their normal states (Fig. 2b), we believe that the transport properties measured in these devices reliably reflect the bulk characteristics of NbGe$_2$ in the normal state. For instance, one of the chirality-induced anomalies in NbGe$_2$ is the coupled electron-phonon liquid [21], leading to a kink feature in the Kohler scaling [30]. We also measured the magnetoresistance and indeed observed similar results in NbGe$_2$ device D1 (Supplementary Fig. 1), demonstrating the high quality of our NbGe$_2$ devices. According to the crystal symmetry, the chirality-induced nonreciprocal transport response $R = R_0[1 + \gamma(\bm{B} \cdot \bm{I})]$ appears when $I$ and $B$ are applied along the helix axis [1,31]. And the nonreciprocal coefficient $\gamma = \frac{2R^{2\omega}}{R^{\omega}BI}$ is obtained by the associated first-harmonic resistance $R^{\omega}$ and second-harmonic resistance $R^{2\omega}$ in lock-in measurements [32]. Figs. 2c and 2d show the raw data of field dependent $R^{2\omega}/R^{\omega}$ measured under red $I//B//c$ and black $I//a \perp B//c$ configurations, respectively. The $R^{2\omega}/R^{\omega}$ data consists of extrinsic symmetric contribution and intrinsic antisymmetric signals. We then extract the intrinsic $R^{2\omega}/R^{\omega}$ response from the asymmetric part. The $c$-axis nonreciprocal response ($I//B//c$) is significantly stronger than the $a$-axis one ($I//a$, $B//c$) as anticipated (Fig. 2e). With decreasing temperature, the chirality-induced nonreciprocal coefficient $\gamma$ eventually saturates at approximately $7.0 \times 10^{-2}$ T$^{-1}$ A$^{-1}$ at low temperatures (Fig. 2f), which is of the same order as that observed in the chiral molecular conductor [33]. In principle, the two-fold rotational symmetry ($2_{001}$) prohibits the generation of nonreciprocal response in any direction other than along the $c$-axis. The observation of a tiny $R^{2\omega}$ response along the $a$-axis ($I//a \perp B//c$) suggests the presence of additional symmetry breaking in the NbGe$_2$ device beyond its intrinsic crystal symmetry, potentially arising from the polar symmetry along the $b$-axis of the artificial surface, as discussed below.

**Type-II superconductivity in artificial NbGe surface**

It is well known that the FIB damages the material surface, leaving an artificial amorphous layer due to ion irradiation [24]. The FIB-induced surface damage is gentle and confined to a typical thickness of around 10 nm, ensuring that the transport properties remain unaffected in FIB-fabricated microdevices, as confirmed by our measurement in the normal state (Fig. 2b). However, the situation became complicated when we investigated the superconducting state. As shown in Fig. 3, we found that both $T_c$ and upper critical fields $H_{c2}$ are significantly enhanced from the NbGe$_2$ bulk ($T_c$ = 2.01 K, $H_{c2}$ = 40 mT) to NbGe$_2$ devices ($T_c$ = 2.95 K, $H_{c2}$ = 4 T in D1). The enhanced superconductivity originates from the amorphous NbGe surfaces induced by FIB irradiation, as demonstrated by the following observations. Firstly, the scanning transmission electron microscope (STEM) image provides direct evidence of an amorphous surface, with a thickness ranging from 10 to 15 nm (Fig. 3b). The six-fold symmetry is observed in the crystalline region 2 and absent in the amorphous region 1 as expected. The composition of the amorphous surface is identified as NbGe using spatial-resolved electron energy loss spectroscopy (EELS) (Supplementary Fig. 2). Secondly, the broad superconducting resistive transition is fitted well by the Halperin-Nelson formula [34], showing a Berezinskii-Kosterlitz-Thouless (BKT) transition with a higher $T_c$ = 2.95 K in device D1 (Fig. 3c). The fitted BKT transition temperature $T_{BKT}$ in the resistivity is 2.50 K, which is almost identical to $T_{BKT}$ = 2.49 K derived from the fitted exponent $V \propto I^\alpha$ at $\alpha(T_{BKT})$ = 3 (Fig. 3d). Note that the BKT transitions were reproducibly observed in another NbGe$_2$ device D2 with a different $T_c$ = 2.50 K (Supplementary Figs. 3a-d). Lastly, the upper critical fields $H_{c2}$ of NbGe$_2$ device are two orders of magnitude higher than that of NbGe$_2$ bulk (Figs. 3e-f), but are consistent with the $H_{c2}$ of NbGe [35] (Supplementary note 3). It is anomalous to observe such a considerable enhancement of $H_{c2}$ if the superconductivity comes from NbGe$_2$ bulk.

Based on these results, the amorphous NbGe shell as depicted in Fig. 3a results in enhanced superconducting properties, such as $T_c$ and $H_{c2}$, in transport measurements. Similar amorphous niobium surfaces were created and responsible for the induced superconductivity in non-superconducting NbAs devices via FIB milling [36]. Since the double superconducting coherence length $2\xi$ = 32 nm is larger than the thickness of NbGe surface (see Supplementary Note 4), our results strongly suggest that the NbGe surface hosts 2D superconductivity in NbGe$_2$ devices [37,38]. Additionally, it is worth noting that the vortex dynamics is always absent (strongly suppressed) below the critical field $H_c$ ($H_{c2}$) of NbGe$_2$ due to its type-I (-II/1) superconductivity and its significantly larger cross section compared to the NbGe surface. However, as confirmed by the observations of BKT transitions, the excitations of vortices/antivortices could penetrate the devices due to the type-II nature of the amorphous NbGe surfaces at higher temperature. This result indicates that we can easily switch between type-I and type-II superconductivity in different regimes, as delineated in the $H$-$T$ phase diagram of Fig. 3f.

**Nonreciprocal transport near $T_c$ in NbGe$_2$ devices**

The amorphous NbGe surface is distinctly separated from the crystalline NbGe$_2$ core, resulting in a well-defined NbGe$_2$/NbGe interface. This sharp boundary is attributed to the self-annealing process occurring at room temperature [24]. Akin to other artificial interfaces, this interface will induce the Rashba-like SOC, which gives rise to polarization in the perpendicular direction. To investigate this additional space-inversion symmetry breaking, we systematically investigated the nonreciprocal electrical transport in NbGe$_2$ devices near $T_c$. The polarization-induced second harmonic resistance $R^{2\omega}$ is expressed as $R^{2\omega} \propto (\boldsymbol{P} \times \boldsymbol{B}) \cdot \boldsymbol{I} = -BI\sin\theta$, where $\theta$ represents the angle between the current $I$ and magnetic field $B$ when they are perpendicular to the polarization $P$ [22,39]. We indeed observed significant $R^{2\omega}$ responses in NbGe$_2$ device D2 under the polar configuration of $I \perp B \perp P$ (Fig. 4a) and its typical angular dependence (Fig. 4b). Fig. 4c shows the temperature dependence of $\gamma$, which features a peak at $T_{BKT}$ with a value of $3.2 \times 10^5$ T$^{-1}$ A$^{-1}$. Fig. 4d illustrates that $R^{2\omega}$ progressively deviates from the anticipated linear behavior in the low-current regime and exhibits a decline in the high-current regime, consistent with previous observations in gated SrTiO$_3$ [40]. Figs. 4e and f show the magnetic field dependence of first-harmonic resistance $R^{\omega}$ and $R^{2\omega}$ at various $T$ down to 0.3 K respectively. It shows that the $R^{2\omega}$ responses are noticeable in the superconducting transition regions but undetectable in the normal state. All the current, temperature, and field dependencies of $R^{2\omega}$ are well reproducible in another NbGe$_2$ device D1 (Supplementary Fig. 4).

These observations confirm the presence of a polar NbGe$_2$/NbGe interface, which induces a significant nonreciprocal transport in NbGe$_2$ devices near $T_c$. There are two major factors that contribute to the substantial increase of $\gamma$ from around $10^{-1}$ T$^{-1}$ A$^{-1}$ in the normal state to approximately $10^5$ T$^{-1}$ A$^{-1}$ in the superconducting state. First, due to the remarkable reduction of the electronic energy scale from Fermi energy to superconducting gap, the $\gamma$ value is gigantically increased when electrons condense into Cooper pairs [41]. The second factor is the BKT transition, which further enhances the $\gamma$ value in the 2D NbGe$_2$/NbGe interfaces. This involves a two-step evolution of $\gamma(T)$ [41], characterized by $\gamma(T) \sim (T_c - T)$ and $\gamma(T) \sim (T - T_{BKT})^{-3/2}$, as the temperature decreases from $T_c$ to $T_{BKT}$. With a further decreasing $T$, $\gamma$ falls to zero associated with the bound vortex/antivortex pairs in the superconducting ground state. This evolution explains why we observe a peak feature in Fig. 4c. It is worth noting that our observations are highly similar to previous reports in various platforms including gated MoS$_2$ ($\gamma \sim 8 \times 10^2$ T$^{-1}$ A$^{-1}$) [32], gated SrTiO$_3$ ($\gamma \sim 3.2 \times 10^6$ T$^{-1}$ A$^{-1}$) [40], Bi$_2$Te$_3$/FeTe interfaces (2D $\gamma' \sim 6.5 \times 10^{-3}$ T$^{-1}$ A$^{-1}$ m) [42], NbSe$_2$ antennas ($\gamma \sim 3.4 \times 10^4$ T$^{-1}$ A$^{-1}$) [43], etc. Albeit these materials possess various origins of broken space-inversion symmetry, the theoretical analysis indicated that the nonreciprocal responses are similar in these 2D superconductors due to the BKT transition [41]. Our results show that the motion of unpaired vortices boosts $\gamma$ significantly in NbGe$_2$ devices, emphasizing the importance of vortex dynamics in nonreciprocal transport [41,44].

**Absence of SDE in type-I NbGe$_2$ regime**

Albeit resistance, whether nonreciprocal or not, vanishes in the superconducting ground state, the supercurrents retain the nonreciprocal characteristics, as manifested itself by the SDE [3]. Owing to a small penetration depth, the theoretical depairing critical current $I_c$ of NbGe$_2$ is significantly larger than that of the NbGe surface (see Supplementary Note 6). This feature suggests that the superconductivity of NbGe$_2$ can be detected in terms of $I_c$ (Fig. 5a), in contrast to the $H$-$T$ phase diagram (Fig. 3f), where the superconductivity of NbGe$_2$ is obscured by the NbGe surface. Hence, the $I$-$V$ characteristics can be divided into two different regimes based on the $H_{c2}$ of NbGe$_2$: the type-I (or type-II/1) NbGe$_2$ regime at low magnetic fields and the type-II NbGe regime at high magnetic fields.

We characterized the *I-V* curves in both positive and negative directions under different magnetic fields at 0.3 K, and summarized the field dependence of $I_c$ in Fig. 5b. The difference between $I_c^+$ and $I_c^-$ becomes noticeable once $B$ exceeds 40 mT, aligning with the $H_{c2}$ of NbGe$_2$ at 0.3 K. The associated *I-V* curves exhibit clear distinctions between the pale purple NbGe$_2$ regime and the pale orange NbGe regime. In the NbGe$_2$ regime like $B$ = 20 mT, the $I_c$ is equal in opposite directions $I_c^+ = I_c^-$. A tiny nonreciprocity appears in the middle dissipative regime, which is difficult to define as the SDE phenomenon (Fig. 5c). In contrast, when the applied magnetic field increases to 100 mT, $I_c$ becomes distinct along opposite directions, $I_c^+ \neq I_c^-$, showing the emergence of SDE in the NbGe regime (Fig. 5d). One vital functionality of SDE is the perfect rectification effect without dissipation, which is also verified in Supplementary Fig. 5. We notice that its polarity is effectively switched by reversing the magnetic fields, as marked by the contrasting red and gray colors. The switchable polarity of SDE in NbGe$_2$ device is similar to that in polar artificial multilayers [4,45], patterned thin-films [13,15], few-layer NbSe$_2$ [46] and others, but in contrast to the strained trigonal superconductor PbTaSe$_2$ [47], where the polarity of SDE is fixed and the underlying diode mechanism should be distinct from others.

To characterize the difference in vortex dynamics between NbGe$_2$ and NbGe regimes, we extract the vortex creep rate *S* using the relation $S \sim 1/(N-1)$, where the flux creep exponent *N* is derived from *I-V* curves $V(I) \propto I^N$ in the vicinity of $I_c$ [48]. Fig. 5e displays the corresponding field dependence of *S*, in which the slope of *S* abruptly increases at the same threshold of 40 mT. This result confirms that the vortex dynamics are strongly suppressed in the NbGe$_2$ regime but effectively activated in the NbGe regime. Additionally, the structural disorder within the amorphous NbGe layer could broaden the *I-V* transition in the type-II regimes (fig. 5d) compared to the type-I regimes (fig. 5c). Given that two sources of space-inversion symmetry breaking are observed in NbGe$_2$ devices, we measured *I-V* curves under both configurations: $I \perp B \perp P$ (devices D1 and D2) and $I//B//c$ (device D3). The *I-V* characteristics are highly consistent across different NbGe$_2$ devices, and the field dependence of their diode efficiency *Q* are shown in Fig. 5f. Distinct from the linear enhancement of *Q* (*B*) in previous reports of SDE [4,13], the *Q* value fluctuates within 2% below 40 mT, indicating the absence of SDE even under conditions of simultaneous space-inversion and time-reversal symmetry breaking. The similar evolution of *Q*(*B*) and *S*(*B*) suggests an underlying relationship between them, highlighting the significance of vortex dynamics in triggering SDE. We also observed small *Q* values ~ 0.2 when the external magnetic fields were parallel to the currents in device D3, a configuration in which the unidirectional Lorentz force should be absent, and thus the SDE is expected to be negligible. However, the applied currents inevitably generate vortex and antivortex pairs, known as the self-field effect [12]. The excitation of vortex/antivortex pairs could induce giant nonreciprocality even when $B$ = 0 [44], providing a possible explanation for our observations in the *B//I* configuration.

## Conclusions

In summary, we systematically performed nonreciprocal transport measurements on FIB fabricated $NbGe_2$ devices and confirmed the critical role of vortex dynamics in realizing a significant SDE. By inducing artificial NbGe surfaces through FIB irradiation, we observed enhanced superconductivity in these devices. Both types of space-inversion symmetry breaking, chirality in the $NbGe_2$ bulk and polarization at the $NbGe_2$/NbGe interface, were detected in the nonreciprocal transport measurements. Additionally, two-dimensional vortex excitations emerged on the NbGe surfaces as the type-I superconductivity of $NbGe_2$ bulk was suppressed, leading to the BKT transition and a dramatic increase in the nonreciprocal coefficient. The SDE was negligible in the $NbGe_2$ bulk at low magnetic fields but became pronounced in the NbGe surface under high magnetic fields, aligning with the abrupt increase in vortex creep rates.

Although we did not observe observe the SDE in chiral $NbGe_2$, likely due to the weak antisymmetric SOC, we cannot dismiss the possibility of exotic superconductivity induced SDE, such as the finite momentum Cooper pairing [49]. Due to the inevitable excitation of vortices in type-II superconductors, it is challenging to distinguish between contributions to SDE from intrinsic crystal symmetry and those from extrinsic vortex effects. Therefore, it is crucial to exercise caution when attributing the widely observed SDE to the inherent symmetry breaking in superconductivity, rather than to the artificial asymmetry. Additional phase-sensitive experiments, such as those measuring the interference patterns [50] or Little–Parks oscillations [51], are necessary and promising to support the existence of exotic superconductivity in these platforms.

From the perspective of electronic applications, we have confirmed that artificial superconducting surfaces can host the SDE. This functionality can be readily replicated in various quantum materials containing superconducting elements, such as niobium, using FIB technology. Consequently, we envision a significant expansion of the SDE application into a broader range of electronic devices, including potential integration with non-superconducting quantum materials, such as the Weyl semimetal NbAs. Our findings suggest that the unique properties of these materials could be harnessed in conjunction with SDE to develop advanced electronic components with enhanced performance and functionality.

## Methods

### Crystal synthesis and characterizations

$NbGe_2$ single crystals were synthesized by vapor transport using iodine as the transport agent. At first, the polycrystalline $NbGe_2$ samples, prepared via solid-state reaction at a heating temperature of 850°C, was ground into a powder and mixed with iodine in a glove box. The reactants were then sealed under vacuum in a quartz tube and placed in a two-zone furnace, with the hot end maintained at 850°C and the cold end at 800°C. After two weeks, hexagonal crystals with typical sizes of 0.5 mm × 0.5 mm × 0.3 mm were obtained after ultrasonic cleaning in ethanol. Magnetization measurements were performed in a Quantum Design MPMS-XL system. The isothermal magnetization down to 400 mK was collected in an MPMS-3 system equipped with the He3 option. The $NbGe_2$ crystal structures are drawn by VESTA [52].

**Devices fabrication**

All $NbGe_2$ devices were fabricated by a FEI Helios-600i FIB under an acceleration voltage of up to 30 kV with Ga ion sources. Similar to the methods developed by Moll et al. [24], the whole procedure of FIB device preparation can be summarized as four steps.

First step: *preparing the lamella from bulk samples*. The crystallographic orientation of the $NbGe_2$ bulk was checked by the Laue diffraction and mounted in the SEM sample holder. We then cut the lamella with a typical size of 150 × 40 × 2 $\mu m^3$ from bulk like the routine sample preparation for the transmission electron microscope. A high 65 nA was used to dig the grooves, and a smaller current of 2.5 nA was then used to polish the sidewalls into a flat shape under grazing incidence.

Second step: *transferring the lamella to substrate*. Firstly, we place a hundred-micrometer-sized droplet of Araldite epoxy glue on the sapphire substrate with lithographically prepatterned gold contact pads. We vertically picked up the lamellas from crystals and then horizontally, flat sides down, put them onto the droplet using an ex-situ micromanipulator. The continuous and smooth contacting profile between lamella and droplet was built by the capillary forces of epoxy liquid.

Third step: *evaporating gold electrical contacts*. The droplet was first heated at 150 °C for one hour to release the bubble from the epoxy droplet. A brief argon etching was used to clean the surface of the lamellas before the following gold evaporation. By using a homemade shadow mask, typical Ti/Au (5/300 nm) layers were deposited on the region covering lamellas and contact pads. The deposited gold layers serve as the electrical contacts with a typical contact resistance of a few Ohms.

Final step: *shaping the measured region*. We put the substrate with covered lamellas back into the FIB chamber and proceeded with the final cutting. The covered gold layers were carefully removed by FIB milling at 1.6 nA, 5 kV. Then, the overall shape of the measured region and the position of the corresponding contact was cut by FIB at 0.79 nA, 30 kV. Finally, the outer gold layer is selectively cut through to ensure the applied current flows only as designed for transport measurements.

**STEM on amorphous surfaces**

To characterize the FIB induced amorphous damage, STEM specimens were prepared using the same FIB machine with the same milling parameters as those used for $NbGe_2$ devices. Atomically resolved high-angle annular dark-field scanning transmission electron microscopy (HAADF-STEM) and electron energy loss spectroscopy (EELS) were conducted on a JEOL ARM200F transmission electron microscopy (TEM) equipped with a probe-forming spherical-aberration corrector.

**Electrical transport measurements**

The temperature dependent resistance and magnetoresistance were conducted in a Quantum Design PPMS-9 system. The dc *I-V* measurements were collected by two Keithley meters 6221 and 2182. The ac $R^{\omega}$ and $R^{2\omega}$ were obtained by the Stanford SR830 lock-in amplifiers with a frequency of 137.77 Hz, and a phase offset as -90° was set to collect $R^{2\omega}$ in the nonreciprocal transport measurements. To obtain a higher cooling power during the current-voltage measurements at 300 mK, similar transport measurements were performed in a top-loading He-3 refrigerator with a 15 T superconducting magnet.

Note added: We noticed a similar study on FIB fabricated NbGe$_2$ device during the peer review process [53].

## Data availability

The data that support the plots within this paper and other finding of this study are available from the corresponding author upon reasonable request.

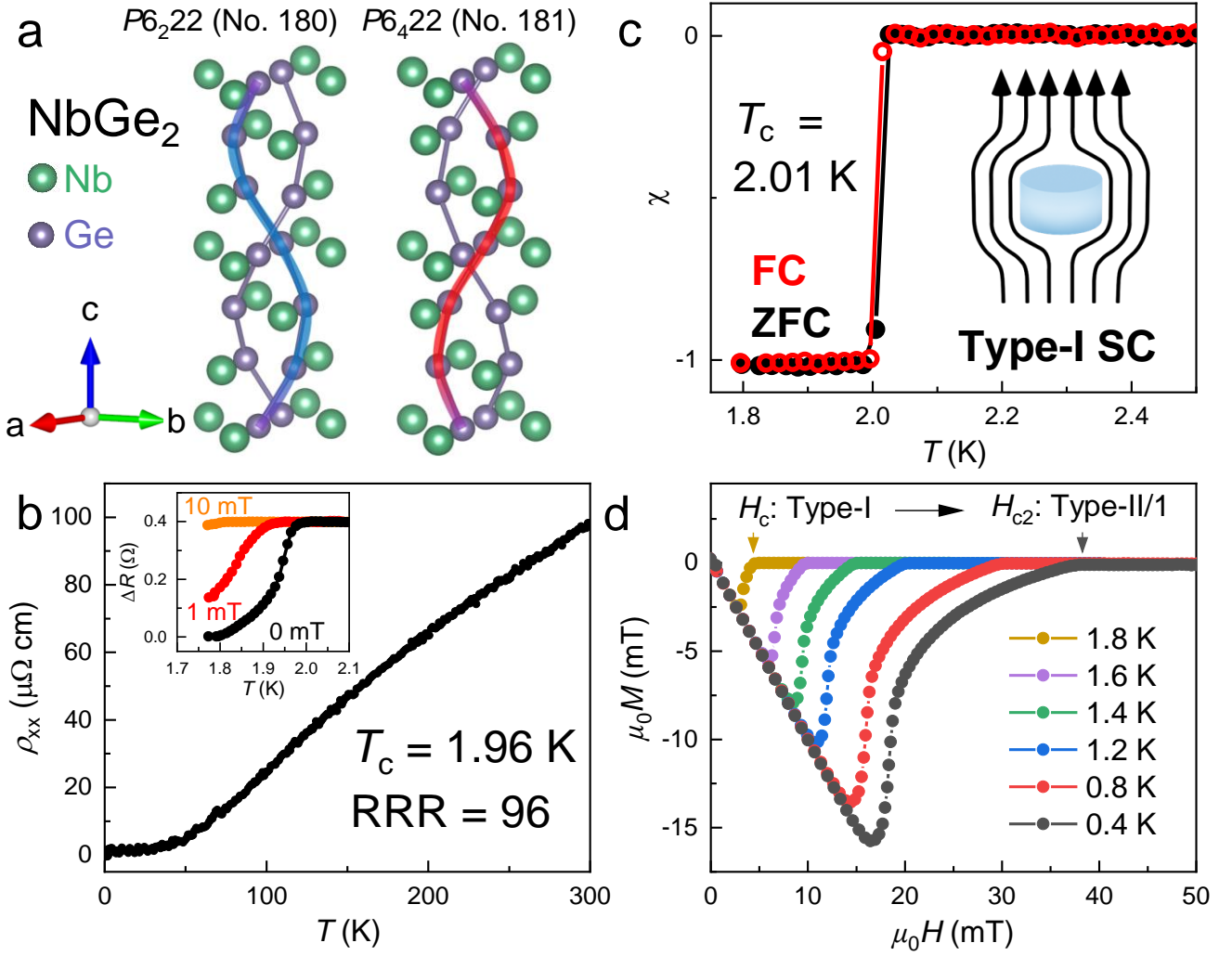

**Fig. 1. Type-I superconductor NbGe$_2$. a**, The chiral crystal structure of NbGe$_2$. The right-handed and left-handed Ge-Ge bond enantiomers are highlighted by blue and red lines, respectively. **b**, The temperature dependent resistivity of a NbGe$_2$ crystal up to 300 K. The residual resistivity ratio RRR = $R_{300 K}/R_{5 K}$ = 96. The inset shows a two-terminal resistance under different magnetic fields, in which the superconducting critical temperature $T_c$ is determined as 1.96 K from the onset of the resistance drop. **c**, The temperature dependent dc magnetic susceptibility $\chi$ of NbGe$_2$ crystal under the zero-field-cooled (ZFC) and field-cooled (FC) modes with an applied magnetic field of $\mu_0 H$= 0.1 mT. $T_c$ is determined as 2.01 K from the onset of diamagnetism. The inset shows the expulsion of the magnetic flux from type-I superconductors. **d**, The isothermal magnetization of NbGe$_2$ crystal from 1.8 K to 400 mK. The data is corrected by the demagnetization factor $N$. The magnetization is expressed as $\mu_0 M$ to align with the external magnetic field $\mu_0 H$.

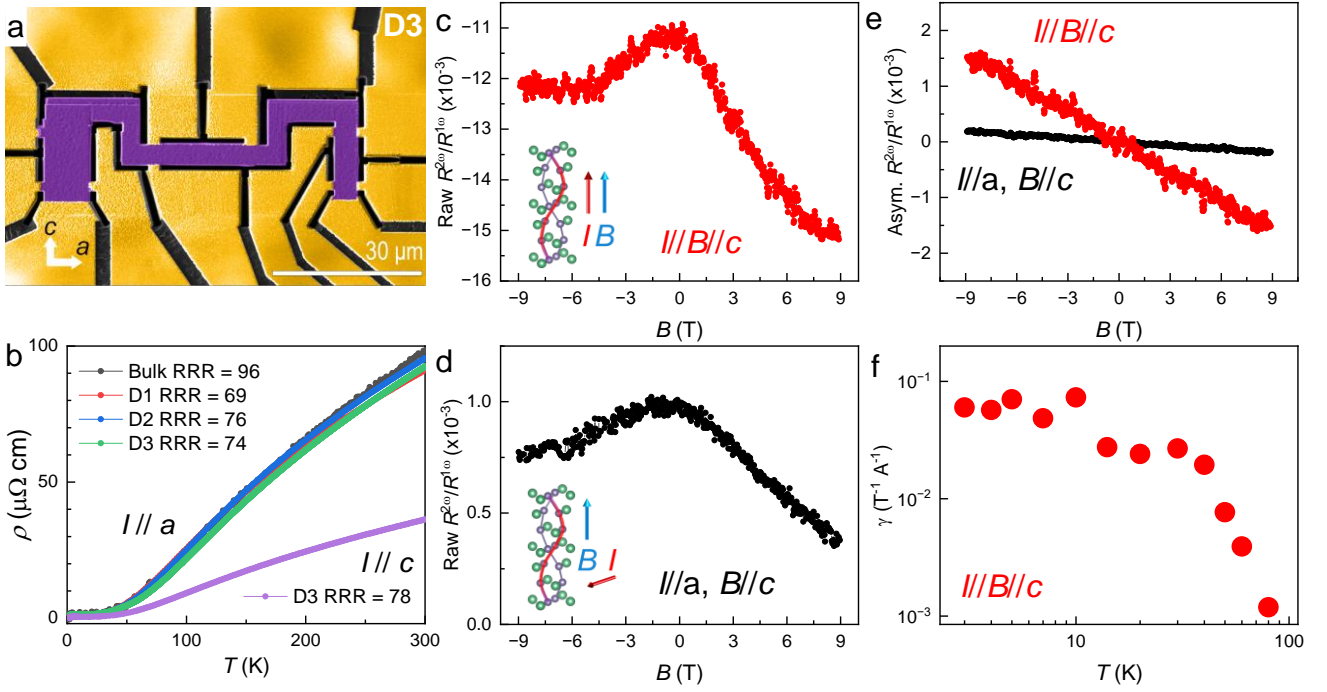

**Fig. 2. Chirality induced nonreciprocal transport in the normal state of NbGe$_2$. a**, Scanning electron microscope image of NbGe$_2$ device D3. The NbGe$_2$ microstructures and electrodes are highlighted in purple and yellow colors, respectively. **b**, The temperature dependent resistivity of NbGe$_2$ devices D1-D3. The purple *c*-axis transport was obtained from D3. The residual resistivity ratio RRR is calculated from $R_{300 K}/R_{5 K}$. **c** and **d**, Raw data of field dependence of second-harmonic resistance $R^{2\omega}$/ first-harmonic resistance $R^{1\omega}$ measured under configurations of current *I*// magnetic field *B*//*c*-axis and $I \perp B$ at $T$ = 10 K, respectively. **e**, Asymmetric analysis of the corresponding raw data shown in a and b. **f**, Temperature dependence of chirality-induced nonreciprocal coefficient $\gamma$ in the normal state. To amplify the subtle $R^{2\omega}$ signal, the applied current was as large as 5 mA. All $R^{2\omega}$ data were obtained on device D3.

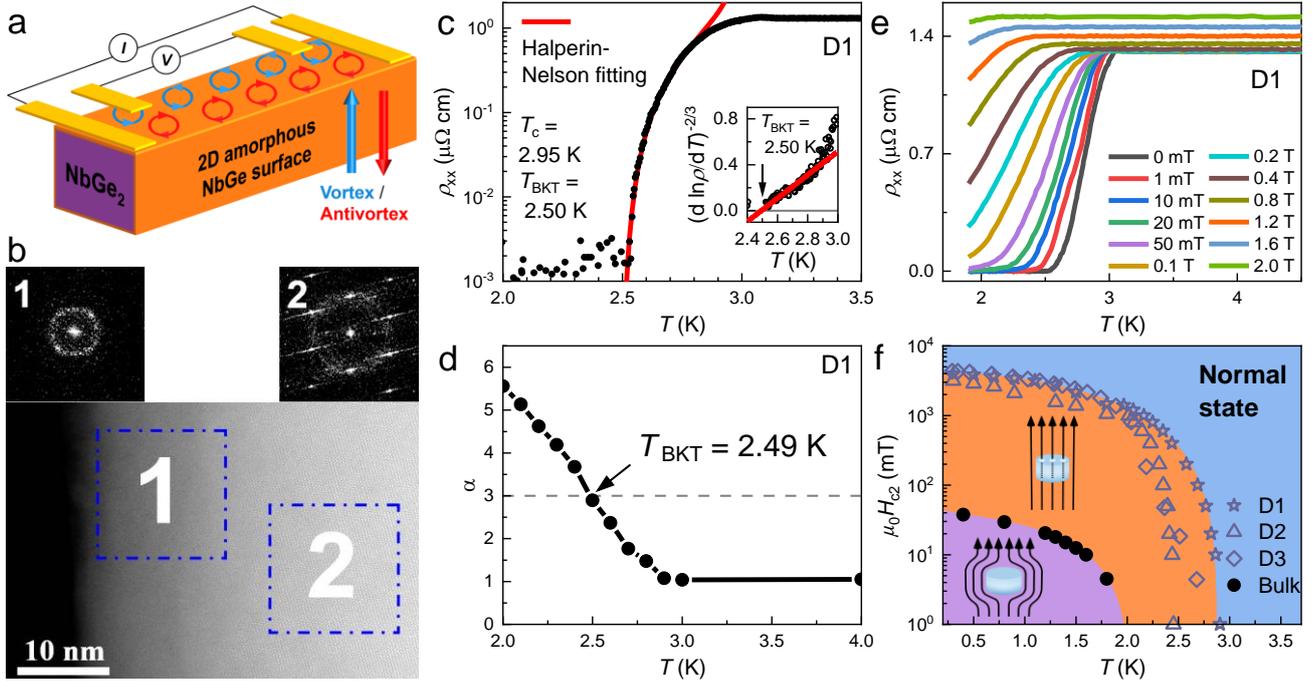

**Fig. 3. Surficial type-II superconductivity in NbGe$_2$ devices. a**, Schematic illustration of NbGe$_2$ device in the transport measurements. The crystalline NbGe$_2$ core is covered by a thin amorphous NbGe shield, which induces the excitations of vortex-antivortex pairs. **b**, High-angle annular dark field (HAADF) image of focused-ion-beam irradiated NbGe$_2$. The corresponding fast Fourier transform (FFT) images of the amorphous region 1 and crystalline region 2 are shown in the top panel. **c**, Detailed scan of resistivity in device D1 with temperature step of 5 mK. The resistive curve is fitted well by the Berezinskii-Kosterlitz-Thouless (BKT) transition using the Halperin-Nelson formula (red line). The superconducting critical temperature $T_c$ is determined at 90% normal state resistivity $\rho_n$ and the fitted $T_{BKT}$ is 2.50 K. The inset is resistivity replotted on a $[d(\ln\rho)/dT]^{-2/3}$ scale. **d**, Temperature dependence of the power-law fitted exponent $\alpha$ from zero-field voltage-current (V-I) curves $V \propto I^\alpha$. The corresponding BKT transition temperature $T_{BKT} \sim 2.49$ K is obtained at $\alpha(T_{BKT}) = 3$. **e**, The resistivity of NbGe$_2$ device D1 as a function of temperature under different magnetic fields. The upper critical field $H_{c2}$ was obtained at 90% $\rho_n$. **f**, Superconducting phase diagram of NbGe$_2$ devices. The solid and open symbols represent the corresponding $H_{c2}$ of NbGe$_2$ bulk and devices D1-D3, respectively. The regime of type-I (type-II/1) superconductivity of the NbGe$_2$ core, type-II superconductivity of the NbGe surface, and the normal state are shown in purple, orange, and blue colors, respectively.

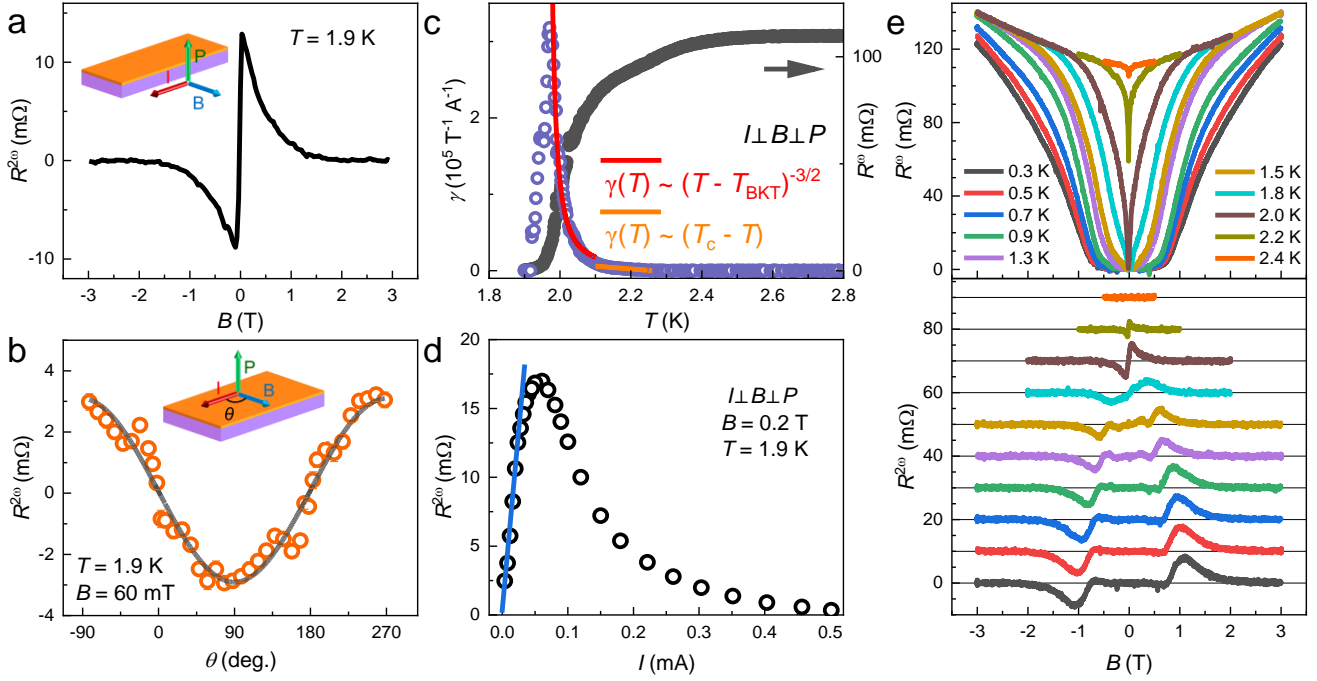

**Fig. 4. Nonreciprocal transport of a NbGe₂ device D2 near the superconducting transition. a**, The field dependence of second-harmonic resistance $R^{2\omega}$ at $T = 1.9$ K under polar configuration. Insets show the interfaces between purple NbGe₂ bulk and orange NbGe surface. The direction of current $I$ (red arrow), external magnetic field $B$ (blue arrow), and polarization $P$ (green arrow) are perpendicular to each other. **b**, Angular dependence of $R^{2\omega}$ measured under 60 mT, 1.9 K. The evolution is fitted well by a gray sine wave. **c**, Temperature dependence of the nonreciprocal coefficient $\gamma$ and first-harmonic resistance $R^\omega$ near the superconducting transition. The red and orange fitting line is Berezinskii-Kosterlitz-Thouless (BKT) transition induced nonreciprocal response $\gamma(T) \propto (T - T_{\mathrm{BKT}})^{-3/2}$ near BKT transition temperature $T_{\mathrm{BKT}}$ and $\gamma(T) \propto (T_c - T)$ near superconducting critical temperature $T_c$. **d**, $R^{2\omega}$ as a function of current $I$ at 1.9 K, 0.2 T. The $R^{2\omega}$ is linear with $I$ in the low current region but then gradually deviates the linearity with an increasing $I$. **e**, The field dependence of $R^\omega$ and $R^{2\omega}$ at various temperatures. The curves of $R^{2\omega}$ are separated vertically by 10 ohms for clarity.

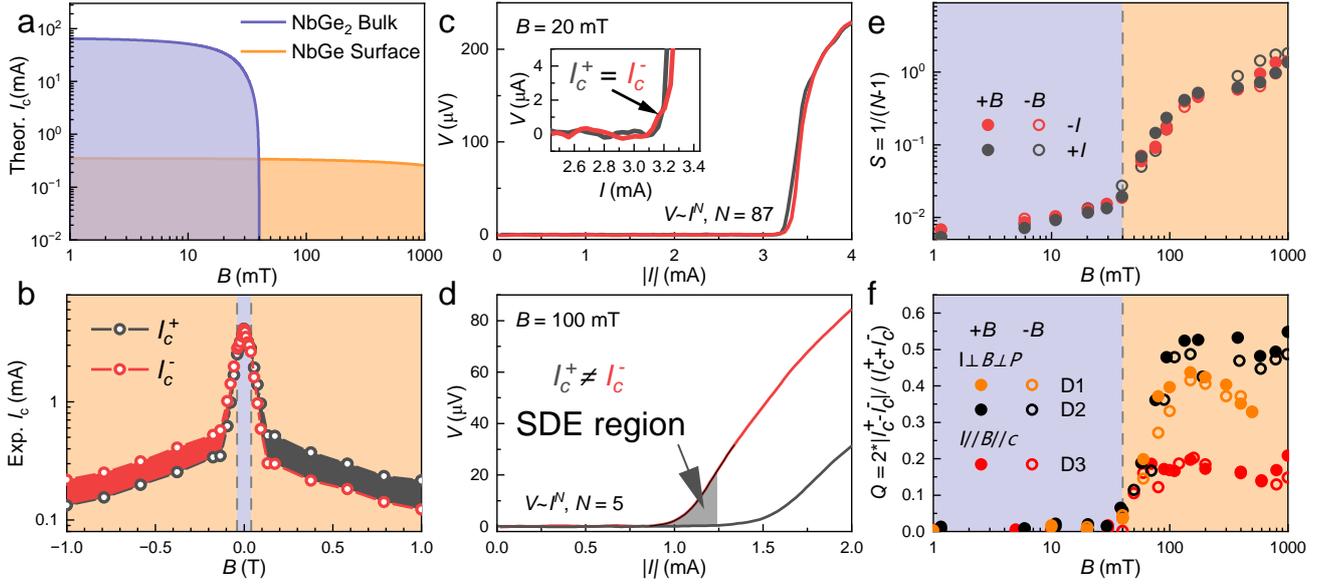

**Fig. 5. Anomalous field dependence of diode efficiency $Q$ in NbGe$_2$ devices. a**, Field dependence of theoretical critical current $I_c$ in NbGe$_2$ devices. $I_c$ maximum is calculated from the self-field $I_c$, and the empirical evolution is capped at the measured upper critical field $H_{c2}$. **b**, The experimental field dependence of $I_c^+$ (positive direction) and $I_c^-$ (negative direction). The low-field NbGe$_2$ and high-field NbGe regimes are marked by pale purple and orange backgrounds. The opposite polarity of SDE is noted by contrasting red and black colors. **c**, and **d**, The typical current-voltage $I$-$V$ curves in low-field NbGe$_2$ and high-field NbGe superconducting regime, respectively. Black and red lines mark the positive and negative current directions. We extract $I_c$ values using the same voltage criterion of 1 μV. The inset of **c** shows an equal critical current $I_c$ along opposite directions. The SDE was only observed in high-field regime, as marked the gray region in **d**. **e**, The vortex creep rate $S$ as a function of fields. The slope of $S$ changes abruptly as superconductivity transitions from the NbGe$_2$ to NbGe superconducting regime. The solid and open symbols present the data obtained from positive and negative magnetic fields, respectively. **f**, The field dependence of $Q$ in NbGe$_2$ devices D1-D3. We measured $I$-$V$ curves under both polar and chiral configurations: current $I \perp$ magnetic field $B \perp$ polarization $P$ (devices D1 and D2) and $I//B//c$-axis (device D3). The $Q$ values are anomalously absent in the NbGe$_2$ superconducting regime.

## Acknowledgments

We acknowledge Junyi Ge, Hiroshi M. Yamamoto, Philip J.W. Moll, and Zhongxian Zhao for their insightful discussions; Max Hirschberger, Ding Zhang, Xiuzhen Yu, Naoto Nagaosa, and Yoshinori Tokura for their useful comments; Aizi Jin and Chunyu Guo for their assistance in preparing FIB devices. We thank Mari Ishida for the assistance in depicting schematic figures. This work was supported by the National Key Research and Development Program of China (Grant Nos. 2022YFA1602803, 2022YFA1403903, and 2023YFA1406100), the National Natural Science Foundation of China (Grant Nos. 92065203, 11527806, 12374141, U22A6005), the Strategic Priority Research Program of Chinese Academy of Sciences (XDB33010200, XDB33010300, XDB33000000), the Innovation Program for Quantum Science and Technology (Grant No. 2021ZD03026001), the CAS Project for Young Scientists in Basic Research (No. 2022YSBR-048), the Beijing Nova Program of Science and Technology (20220484014). A portion of this work was carried out at the Synergetic Extreme Condition User Facility (SECUF).

## Author contributions

D. Li conceived this project. Crystals were synthesized and characterized by Z.-Y.W. L., D.L., Y.L., T.S., X.-L.D., and J.-G. C. The FIB devices were prepared by D.L., W.-X.C., X.-F.S., and J. S. The electrical transport measurements were conducted by D.L., L.-H.H., W.-X.C., T.S., P.-L.L., H.Z., Z.-Y.L., Q.-H.C., K.J., F.-M.Q., and G.-T.L. TEM experiments were performed and analyzed by X.-P.M. and H.-X. Y. The experimental results were analyzed by D.L., Y.M.I., J. H., and Y. I. D.L., Q.-H.C., and G.-T.L. wrote the manuscript with input from all authors. X.-L.D., K.J., J.-G. C., H.-X. Y, Li Lu, J.-P. Hu, and Y. I. coordinated this research project.

## Competing interests

The authors declare no competing interests.

## Correspondence and requests for materials should be addressed to

Dong Li, Qihong Chen, or Guangtong Liu.

# Supplementary Materials

## Supplementary Note 1: Structure refinement of NbGe$_2$ crystal

The crystallographic parameters for NbGe$_2$ are summarized in Supplementary Table S1. A specimen was used for X-ray structural refinement analysis using a Bruker D8 VENTURE diffractometer. X-ray intensity data were measured ($\lambda$ = 0.71073 Å) with a total exposure time of 6 hours. The frames were integrated with the Bruker SAINT software package employing a narrow-frame algorithm. The data set was corrected for absorption effects using the multi-scan method (SADABS). The ratio of minimum to maximum apparent transmission was 0.635. The structure was solved and refined with the Bruker SHELXTL Software Package (Sheldrick, 2018) in the space group $P6_422$, with Z = 3 for the formula unit NbGe$_2$. The final anisotropic full-matrix least-squares refinement on $F^2$ with 9 variables converged at $R_1$ = 3.24% for the observed data and $wR_2$ = 7.03% for all data. The goodness-of-fit was 1.140. The isotropic displacement parameters were 0.0028(4) and 0.0053(4) Å$^2$ for Nb and Ge, respectively.

Supplementary Table 1: **Single crystal X-ray diffraction refinement parameters**.

| Formula | NbGe$_2$ |
| --- | --- |
| Formula weight (g/mol) | 238.09 |
| Space group | $P6_422$ (No. 181) |
| Lattice parameters (Å) | a = b = 4.9653(5) |
|  | c = 6.7814(7) |
| Unit cell volume (Å$^3$) | 144.79(3) |
| Z | 3 |
| Density (g/cm$^3$) | 8.192 |
| $\Theta$ range | 4.74 - 28.22 |
| Reflections collected | 2261 |
| Independent reflections | 128 |
| Absolute structure parameter | 0.10(9) |
| Goodness-of-fit on F$^2$ | 1.140 |
| $R_1[F^2>2\ \sigma(F^2)]$ | 0.0324 |
| $wR_2(F^2)$ | 0.0703 |
| Largest diff. peak and hole (e/ Å$^3$) | 0.860 and -1.370 |

## Supplementary Note 2: Evidence of coupled electron-phonon liquid in NbGe₂

The electron hydrodynamics phenomenon was recently found in ultraclean materials, such as $PdCoO_2$ [54] and graphene [55], opening a new area of investigation. Due to the ultralow residual resistivity, $NbGe_2$ emerges as a promising candidate material for studying electron hydrodynamics. Supplementary Fig. 1 shows evidence of a coupled electron-phonon liquid in $NbGe_2$. To enhance the resistance significantly, we intentionally fabricated the $NbGe_2$ device D1, which is 100 μm long and 6 μm wide (Supplementary Fig. 1). The chiral structure of $NbGe_2$ facilitates the emergence of this coupled electron-phonon liquid in $NbGe_2$ below a certain temperature $T'$ [21]. Indeed, the Kohler scaling shows a change of slope around $T' \sim 50$ K, which is attributed to the transition from the momentum-relaxing Umklapp scattering at high temperatures to a momentum-conserving electron-phonon scattering regime. While the MR of the previously reported $NbGe_2$ bulk (Supplementary Fig. 1d) is higher than that of our sample due to reduced impurity scattering, both exhibit a consistent kink behavior in Kohler scaling around $T' \sim 50$ K.

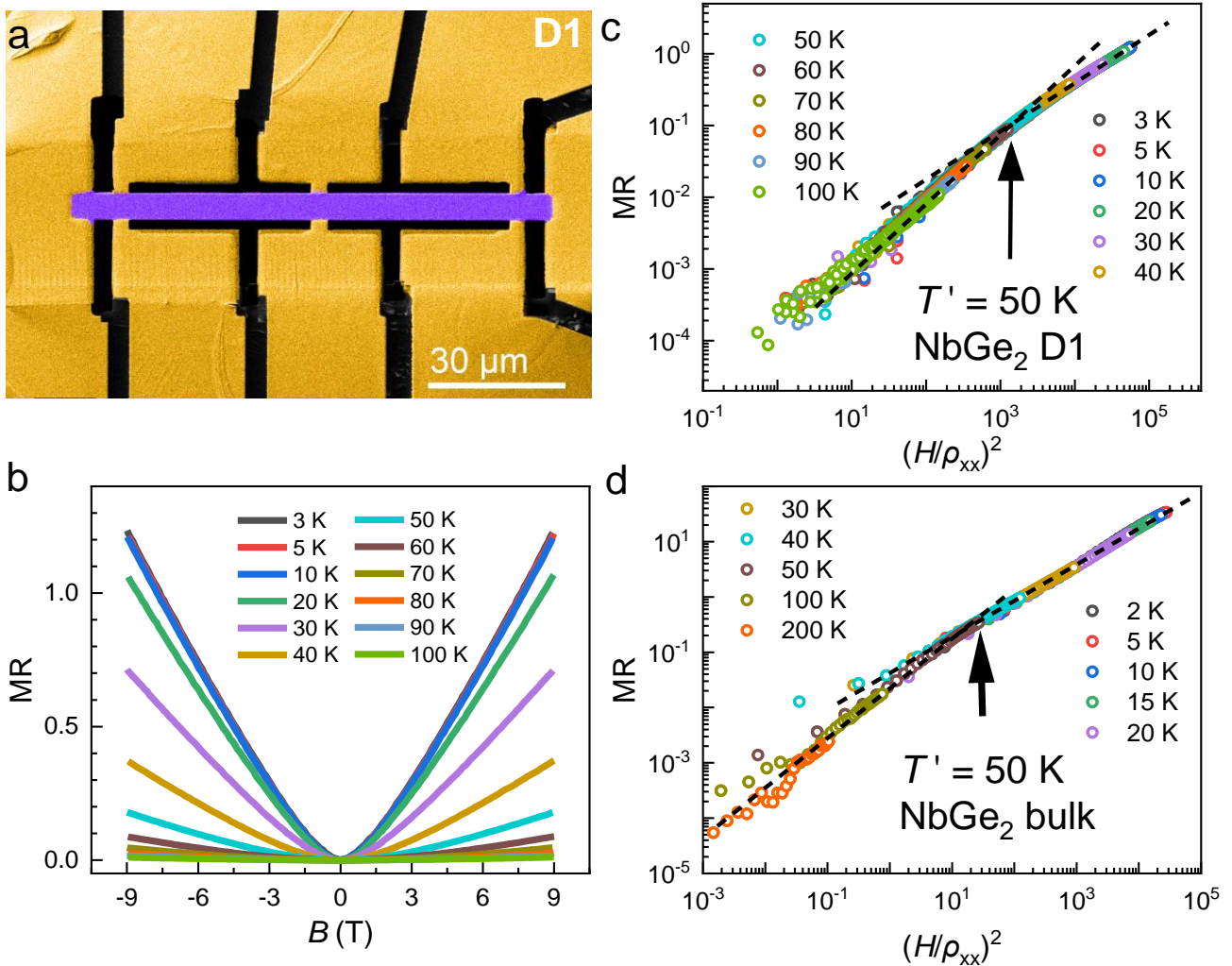

**Supplementary Fig. 1. Chirality-induced coupled electron-phonon liquid in NbGe₂. a.** Scanning electron microscope (SEM) image of $NbGe_2$ device D1. The $NbGe_2$ microstructure and gold electrodes are highlighted in purple and yellow. The resistance is measured along *a*-axis direction. **b.** The trend of magnetoresistance MR = [R(H)

– $R_0$]/$R_0$ from 3 K to 100 K. **c**. The corresponding Kohler scaling analysis. The horizontal axis is scaled in $H/R_{xx}$, where $R_{xx}$ is the longitudinal resistance under $H = 0$. **d**. The Kohler scaling analysis of NbGe$_2$ bulk, which is extracted from ref. [21]. A similar kink behavior shows around $T' \sim 50$ K.

# Supplementary Note 3: The composition of the amorphous layer

The kinetic energy of FIB, produced by 30 kV $Ga^{2+}$ ions, is fully absorbed within the working region with a typical effective depth of 10-20 nm. This process inevitably results in amorphous surfaces in FIB-prepared structures. The milling rate differs for elemental Nb and Ge due to their distinct surface binding energy, leading to an enrichment of niobium at the amorphous surfaces. To characterize this amorphous region, we use the spatially resolved electron energy loss spectroscopy (EELS) to characterize the amorphous region (Supplementary Fig. 2). The EELS signal for Ge exhibits an abrupt decline at the boundary between the crystalline and the amorphous region. Therefore, the composition of the amorphous region is $NbGe_x$ with $x$ gradually decreasing from the core to the surface. For clarity, we refer to this as the amorphous NbGe surface in the main text.

To confirm that the superconductivity comes from the amorphous NbGe rather than the amorphous Nb surface, we compared the upper critical field $H_{c2}$ of $NbGe_2$ device D1, amorphous NbGe, and amorphous Nb, as shown in Supplementary Fig. 2c. The results clearly demonstrate that the enhanced superconductivity in our $NbGe_2$ devices is attributable to the amorphous NbGe.

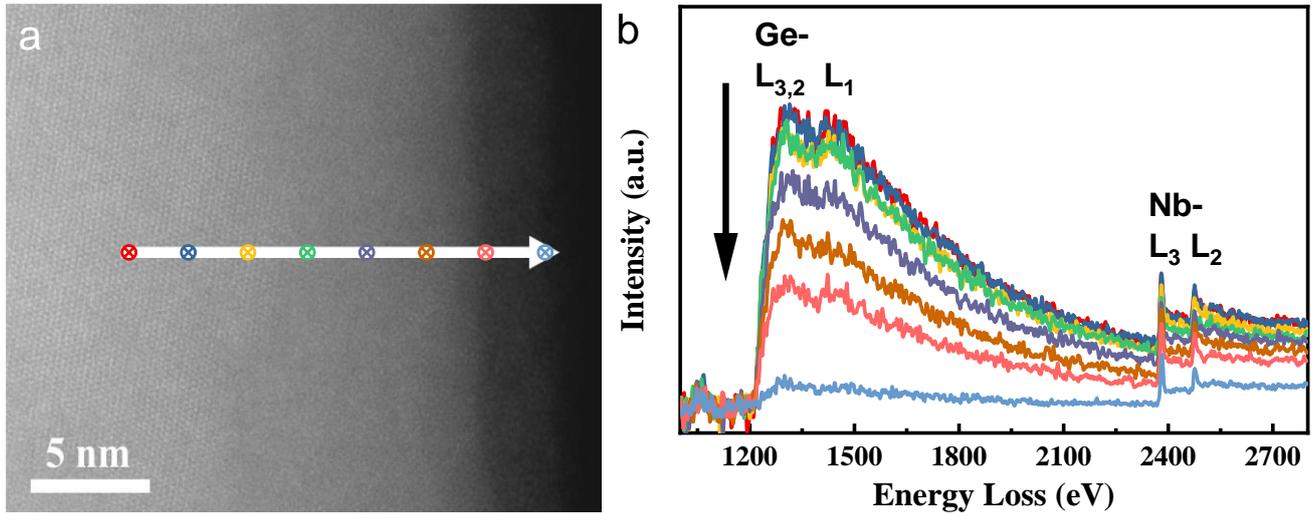

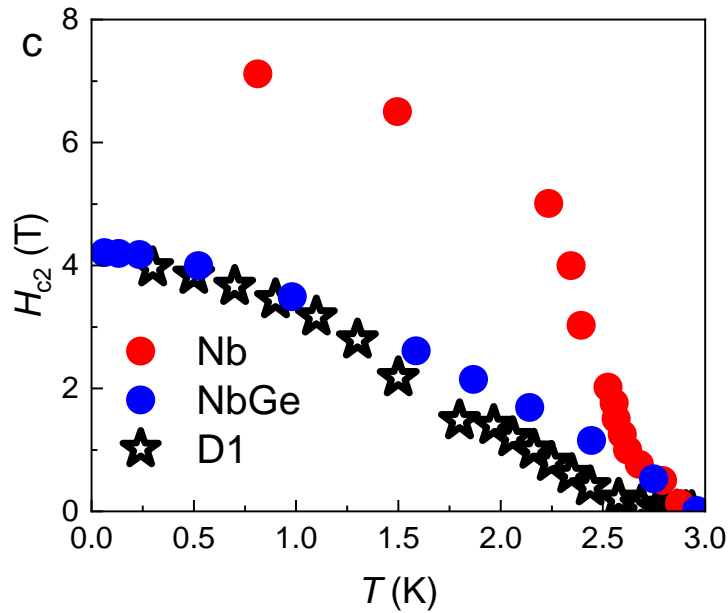

**Supplementary Fig. 2. The composition of the FIB-induced amorphous layer. a**, HAADF image taken near the amorphous surface. The white arrow indicates the collecting direction of EELS. The amorphous region begins to appear around the green spot. **b**, The corresponding EELS data from the crystalline $NbGe_2$ region to the amorphous NbGe surface. The L edges of elemental Ge and Nb are marked respectively. **c**, The temperature dependence of $H_{c2}$ of device D1, Nb [36], and *a*-NbGe [56] shown in different colors.

## Supplementary Note 4: Reproducible 2D superconductivity in NbGe$_2$ devices

We measured another two NbGe$_2$ devices, D2 and D3, to confirm the reproducibility of 2D superconductivity, as shown in Supplementary Fig. 3. The BKT transitions were observed in the device D2 with a $T_c$ of 2.50 K. The fitted BKT transition temperature $T_{BKT}$ from the resistivity data is 2.05 K, almost the same as $T_{BKT}$ = 2.06 K obtained from the $I$-$V$ curves at $\alpha(T_{BKT})$ = 3. It is important to note that the FIB milling process varies across different lattice planes of the same material [57], as implemented in the device D2. This variability results in a lower $T_c$ for device D2 compared to device D1. The $T_c$ of amorphous NbGe is highly sensitive to the Nb/Ge atomic ratio [35], which should be sample dependent. Supplementary Figs. 3e-f shows the temperature dependence of $H_{c2}$ of device D3 when magnetic field is applied along the NbGe surface. We note that the coherence length $\xi$ is roughly estimated as 8.7 nm according to the relationship $\xi = \sqrt{\frac{\Phi_0}{2\pi\mu_0 H_{c2}}}$ in three dimensional superconductors using measured $H_{c2}$ = 4.3 T. When the sample thickness $t$ is smaller than $2\xi$, like the case of NbGe surface $t$ = 15 nm < $2\xi$ = 17.4 nm, 2D superconductivity is realized, as out-of-plane vortex penetration is effectively prohibited in NbGe surface. This result explains why the temperature dependence of $H_{c2}$ is well fitted by the 2D Ginzburg-Landau equation $H_{c2}(T) = \frac{\Phi_0 \sqrt{12}}{2\pi\mu_0 t \xi_c}\sqrt{1-\frac{T}{T_c}}$ in device D3, yielding a more reasonable out-of-plane coherence length $\xi_c$ = 16 nm of NbGe.

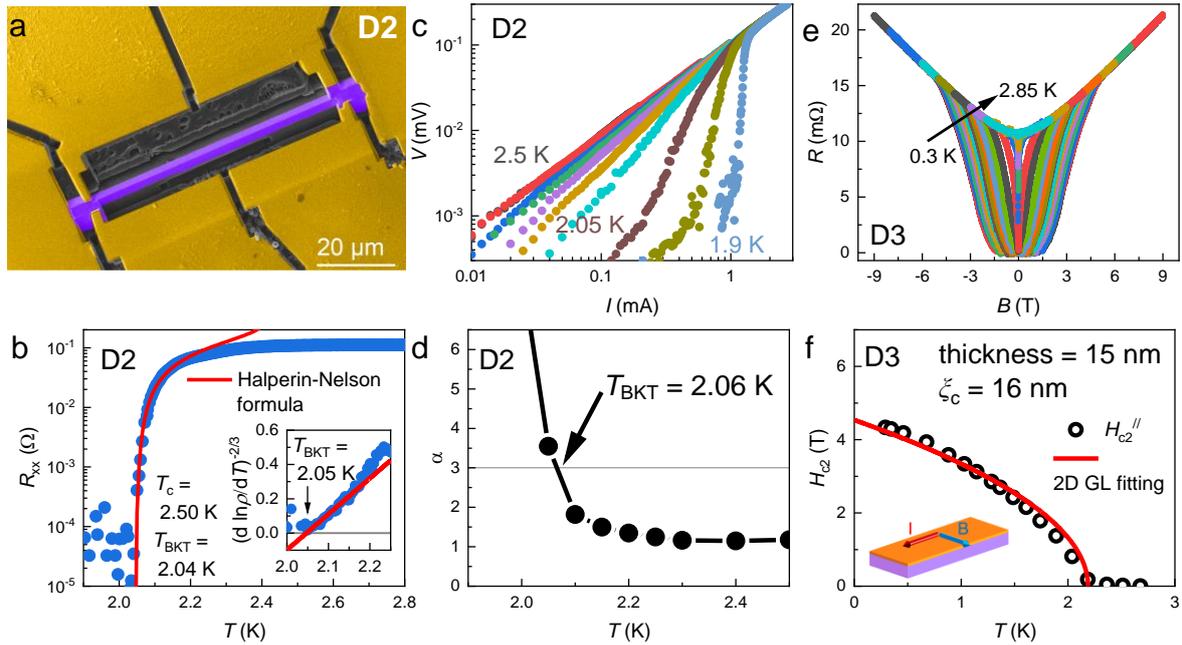

**Supplementary Fig. 3. Reproducible 2D superconductivity in NbGe$_2$ devices. a**, SEM image of NbGe$_2$ device D2 with the same convention. The resistance is measured along $a$-axis direction. **b**, Resistance of device D2 with a $T_c$ = 2.50 K. The resistive curve is fitted well by the BKT transition using the Halperin-Nelson formula (red line). The inset is resistivity replotted on a [d(ln $\rho$)/d$T$]$^{-2/3}$ scale. **c**, $I$-$V$ curves at different temperatures under zero magnetic field of D2. **d**, Temperature dependence of the power-law fitted exponent $\alpha$ from voltage-current curves. **e**, Field dependent resistance of device D3 under various temperatures from 0.3 K to 2.85 K. The magnetic field was applied along the NbGe surface. **f**, The temperature dependence of $H_{c2}^{//}$ of device D3. The in-plane upper critical field $H_{c2}^{//}$ was obtained at 90% $R_n$ in e. The red fitting line is the 2D Ginzburg-Landau equation.

## Supplementary Note 5: Reproducible nonreciprocal transport in NbGe$_2$ devices

Supplementary Fig. 4 shows the reproducibility of nonreciprocal transport in NbGe$_2$ device D1. Due to the absence of nonreciprocal signals in the superconducting ground state at 0.3 K, the linear regime of current dependent $R^{2\omega}$ does not pass the origin (Supplementary Fig. 4a). The temperature dependence of the nonreciprocal coefficient $\gamma$ also shows a peak feature in device D1 (Supplementary Fig. 4b). Its maximum reaches $6 \times 10^5$ T$^{-1}$ A$^{-1}$, which is of the same order as the largest $\gamma$ value in device D2. The field dependence of $R^{2\omega}$ also demonstrates that the nonreciprocal transport is significantly enhanced by the motion of the two-dimensional vortices in device D1 (Supplementary Figs. 4c and d).

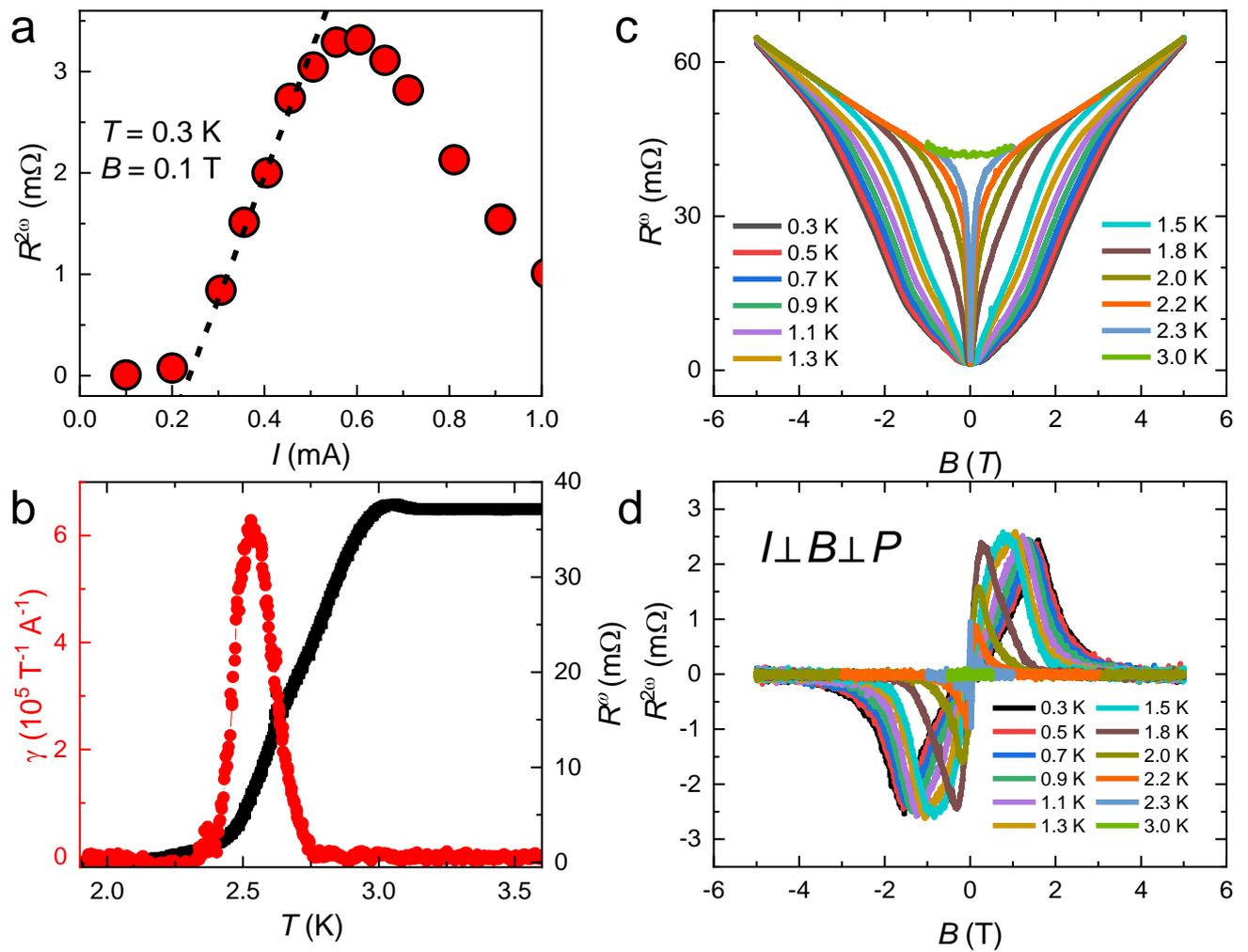

**Supplementary Fig. 4. Reproducible nonreciprocal transport in NbGe$_2$ devices. a**, $R^{2\omega}$ as a function of current $I$ taken on device D1. The $R^{2\omega}$ is only linear with $I$ in the middle current regime. **b**, The temperature dependence of $\gamma$ and $R^{\omega}$ under B = 10 mT. **c** and **d**, The field dependent $R^{\omega}$ and $R^{2\omega}$ at temperature from 0.3 K to 3.0 K, respectively.

## Supplementary Note 6: Estimation of theoretical $I_c(H)$

It is challenging to predict the precise evolution of $I_c(H)$, which is affected by the sample-dependent vortex pinning conditions. As an alternative approach, we can first estimate $J_c(H=0)$, which is known as the self-field critical current density, $J_c(\text{sf})$. The $J_c(\text{sf})$ can be estimated by penetration depth $\lambda$ and superconducting coherence length $\xi$ using the following equations [58]:

$$J_c^{\text{I}}(\text{sf}) = \frac{\Phi_0}{2\sqrt{2}\pi\mu_0\lambda^2\xi} \quad \text{for type-I superconductors} \quad (1)$$

$$J_c^{\text{II}}(\text{sf}) = \frac{\Phi_0}{4\pi\mu_0} \cdot \frac{\ln(\frac{\lambda}{\xi})+0.5}{\lambda^3} \quad \text{for type-II superconductors} \quad (2)$$

Where $\Phi_0$ is the magnetic flux quantum and $\mu_0$ is the magnetic permeability of free space. The parameters $\lambda \sim 125$ nm and $\xi \sim 93$ nm are deduced from the measured $H_c \sim 20$ mT and the $H_{c2} \sim 38$ mT of $NbGe_2$ crystal at 0.4 K. Given that $NbGe_2$ hosts type-II/1 superconductivity at 300 mK, the calculated $J_c(\text{sf})$ is 5.3 MA/cm$^2$ for $NbGe_2$ based on equation (2). Using the parameters $\lambda \sim 620$ nm of amorphous NbGe [59] and $\xi \sim 16$ nm from $H_{c2}$ measurements, the associated $J_c(\text{sf})$ is as 0.23 MA/cm$^2$.

Assuming the cross-section size 0.25 (double $\lambda$) ×5 (width) μm$^2$ for $NbGe_2$ and 0.03 (double $t$) ×5 (width) μm$^2$ for amorphous NbGe surface, the estimated $I_c(\text{sf})$ are 66 mA and 0.35 mA, respectively. This result indicates that the low field $I_c$ is dominated by the $NbGe_2$ bulk rather than the NbGe surface.

# Supplementary Note 7: Rectification effect of superconducting diodes

Supplementary Fig. 5 shows the test of rectification effect in NbGe$_2$ device. In the SDE region, rectified unidirectional resistance occurs in one direction, while superconducting zero resistance is present in the opposite direction. This property is advantageous as it significantly reduces energy dissipation compared to semiconductor p-n junctions. Meanwhile, the polarity of the rectification effect is reversed by switching the magnetic fields.

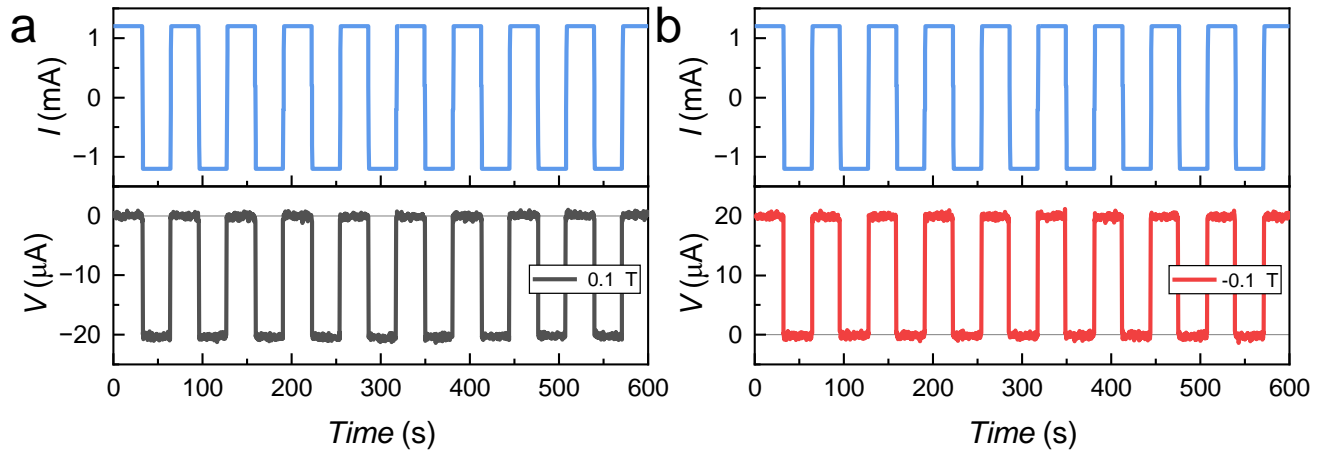

**Supplementary Fig. 5. Rectification effect in NbGe$_2$ device. a**, and **b**, Rectification effect of NbGe$_2$ device D2 at 0.3 K, ±0.1 T, respectively.

# Supplementary References